\documentclass[12pt, a4paper]{article}

\usepackage{rotating}
\usepackage{amsmath}
\usepackage{mathtools}
\usepackage{amsthm}
\usepackage{amsfonts}
\usepackage[dvips]{epsfig}
\usepackage{graphicx}
\usepackage{psfrag}

\usepackage{amssymb}
\usepackage{cancel}
\usepackage{pstricks} 
\usepackage{lscape}
\usepackage{cancel}
\usepackage{slashed}
\usepackage{pstricks} 
\usepackage{lscape}
\usepackage{color}
\usepackage{cite}
\usepackage{url}
\usepackage{caption}
\usepackage{bbm}
\usepackage{bm}
\usepackage{subfig}
\usepackage{bbold}
\usepackage{caption}

\newcommand{\beq}{\begin{equation}}
\newcommand{\eeq}{\end{equation}}
\newcommand{\ga}{\lower.7ex\hbox{$\;\stackrel{\textstyle>}{\sim}\;$}}
\newcommand{\la}{\lower.7ex\hbox{$\;\stackrel{\textstyle<}{\sim}\;$}}

\makeatletter
\newcommand{\Cen}[2]{%
  \ifmeasuring@
    #2%
  \else
    \makebox[\ifcase\expandafter #1\maxcolumn@widths\fi]{$\displaystyle#2$}%
  \fi
}
\makeatother

\begin{document}

\begin{flushright}
{\tt KCL-PH-TH/2017-60}, {\tt CERN-PH-TH/2017-248}  \\
{\tt ACT-06-17, MI-TH-1771} \\
{\tt UMN-TH-3706/17, FTPI-MINN-17/22} \\
\end{flushright}

\vspace{1cm}
\begin{center}
{\bf {\LARGE From $R^2$ Gravity to No-Scale Supergravity}}
\end{center}

\vspace{0.05in}

\begin{center}{\large
{\bf John~Ellis}$^{a}$,
{\bf Dimitri~V.~Nanopoulos}$^{b}$ and
{\bf Keith~A.~Olive}$^{c}$
}
\end{center}

\begin{center}
{\em $^a$Theoretical Particle Physics and Cosmology Group, Department of
  Physics, King's~College~London, London WC2R 2LS, United Kingdom;\\
  National Institute of Chemical Physics \& Biophysics, R{\" a}vala 10, 10143 Tallinn, Estonia; \\
Theoretical Physics Department, CERN, CH-1211 Geneva 23,
  Switzerland}\\[0.2cm]
{\em $^b$George P. and Cynthia W. Mitchell Institute for Fundamental
 Physics and Astronomy, Texas A\&M University, College Station, TX
 77843, USA;\\ 
 Astroparticle Physics Group, Houston Advanced Research Center (HARC),
 \\ Mitchell Campus, Woodlands, TX 77381, USA;\\ 
Academy of Athens, Division of Natural Sciences,
Athens 10679, Greece}\\[0.2cm]
{\em $^c$William I. Fine Theoretical Physics Institute, School of
 Physics and Astronomy, University of Minnesota, Minneapolis, MN 55455,
 USA}

\end{center}

\bigskip
\bigskip

\centerline{\bf {\large ABSTRACT}}
\vspace{0.5cm}
{We show that $R^2$ gravity coupled conformally to scalar fields is equivalent to the real bosonic
sector of SU(N,1)/SU(N)$\times$U(1) no-scale supergravity, where the conformal factor can be identified with the K\"ahler potential,
and we review the construction of Starobinsky-like models of inflation within this framework.} 

\vspace{0.5in}

\begin{flushleft}
November 2017
\end{flushleft}
\medskip
\noindent

\newpage

The singularity-free cosmological model which incorporates inflation~\cite{Starobinsky}, and that in which quantum perturbations were first calculated~\cite{MukhChib},
was that based on $R + R^2$ gravity. Remarkably, almost four decades later, the perturbation
spectrum calculated in this pioneering model of
inflation remains in excellent agreement with the growing wealth of measurements of the cosmic microwave
background (CMB) radiation and data on large-scale structure~\cite{planck15}, whereas many more junior models
have fallen by the wayside. 

Since we are unabashed fans of supersymmetry at the TeV scale and
above, we have long advocated supersymmetric models of inflation~\cite{Cries}. Since any cosmological model
must incorporate gravity, we have also long advocated models of inflation based on in the framework
of local supersymmetry, i.e., supergravity \cite{nost}. In particular, for over 30 years we have been advocating 
models of inflation~\cite{GL,KQ,EENOS,otherns} formulated within no-scale supergravity~\cite{no-scale,EKN,LN}, which offers a positive semidefinite
scalar potential and mitigates the $\eta$-problem \cite{GMO} that is the bane of generic supergravity models of
inflation \cite{drt}. This conclusion is unaffected by radiative corrections \cite{deln,GMO}.

The advent of a new generation of CMB data in the last few years encouraged us to return to the
construction of no-scale supergravity models of inflation. Imagine our surprise when we discovered
that a simple model of an inflaton field coupled to SU(1,1)/U(1) no-scale supergravity \cite{EKN1} could yield an
effective scalar potential that is identical to that obtained in the original $R + R^2$ model after a
conformal transformation~\cite{ENO}, a realization that had also been reached in 1987~\cite{Cecotti}, though without making
the connection to cosmological inflation. This convergence between $R + R^2$ gravity and
no-scale supergravity was very intriguing~\cite{DLT}, but the nature of any deeper connection remained obscure.

In this paper we make a simple point that, to our knowledge, has not been made previously in the way described here.
We consider pure $R^2$ gravity supplemented by a set of complex scalar fields $\phi^i$ with
conformal couplings to $R$ of the form $(\sum_{i=1}^{N-1} |\phi^i|^2)R/3$. Within this extended $R^2$ theory, 
we consider a generalization of the
conformal transformation~\cite{WhittStelle} that rewrites $R + R^2$ gravity as minimal $R$ gravity coupled to
a scalar field with a potential of the form
\begin{equation}
V \; = \; \frac{3 M^2}{4 \kappa^2} \left(1 - e^{-\sqrt{\frac{2}{3}} \phi }\right)^2 \, ,
\label{Staro}
\end{equation}
which yields successful 
inflation. The multifield generalization yields a scalar Lagrangian that is identical to that obtained in a SU(N,1)/SU(N)$\times$U(1)
no-scale supergravity model~\cite{EKN} with K\"ahler potential
\begin{equation}
K \; = \; - 3 \ln \left(T + T^\dagger - (\sum_{i=1}^{N-1} |\phi^i|^2)/3 \right) \, ,
\label{noscale}
\end{equation}
when we discard the imaginary parts of the scalar fields. Within this framework,
the inflaton field $\phi$ in (\ref{Staro}) can be identified (up to a suitable
normalization) with either the combination $T + T^\dagger$ or the real part of one of the conformal fields $\phi^i$. 
Moreover, the conformal factor $\Omega$ that transforms the extended $R^2$
theory to the Einstein frame is identical to the K\"ahler potential (\ref{noscale}), up to a
numerical factor: $\Omega = - K/6$. This identification reinforces the connection
between $R^2$ gravity and no-scale supergravity that emerged in~\cite{ENO} (see also~\cite{KLT}),
and was developed further in~\cite{ENO7,EGNO4,FeKR,others,FKP,Moreothers,RS,EGNNO,EGNNO2}.\\

\noindent
{\it $R^2$ Gravity and a De Sitter Universe}\\

As a preliminary to developing this connection, we first review the case of pure $R^2$ gravity, which is
described by the action:
\begin{equation}
{\cal A} \; = \; \frac{1}{2}\int d^4x \sqrt{-g} \alpha R^2 \, ,
\label{pureR2}
\end{equation}
where $\alpha$ is an arbitrary dimensionless constant. The pure $R^2$ theory (\ref{pureR2}) is scale-invariant.
It may be rewritten in the following form, using a Lagrange multiplier field $\Phi$:
\begin{equation}
{\cal A} \; = \; \frac{1}{2}\int d^4x \sqrt{-g} \left( 2 \alpha \Phi R - \alpha\Phi^2 \right) \, .
\label{phiR2}
\end{equation}
Note that the field $\Phi$ has non-canonical mass dimension: $[\Phi] = 2$, not the canonical mass dimension $[\Phi] = 1$.

In order to rewrite the action (\ref{phiR2}) in the Einstein-Hilbert form,
one rescales the metric by introducing a conformal factor $\Omega$, as follows:
\begin{equation}
{\tilde g}_{\mu \nu} \; = \; e^{2\Omega} g_{\mu \nu}  \; = \; \frac{2 \alpha \Phi}{\mu^2} g_{\mu \nu} \, ,
\label{rescaledmetricR2}
\end{equation}
where $\mu$ is a  mass scale to be determined.
The pure $R^2$ action (\ref{pureR2}) then becomes
\beq
{\cal A} \; = \; \frac{1}{2}\int d^4x \sqrt{-\tilde{g}} e^{-4\Omega} \alpha R^2 \, ,
\eeq
where
\beq
R = e^{2\Omega} {\tilde R} + 6 \Box \Omega + 6 \partial^\mu \Omega \partial_\mu \Omega \, ,
\eeq
which implies that
\beq
e^{-4\Omega} \alpha R^2 = e^{-2\Omega} 2 \alpha \Phi {\tilde R} + 12 e^{-4\Omega} \alpha \Phi (\Box \Omega + \partial^\mu \Omega \partial_\mu \Omega)- e^{-4\Omega} \alpha \Phi^2 \, .
\eeq
After rewriting contractions and covariant derivatives in terms of the new metric ${\tilde g}$,
we have 
\beq
e^{-4\Omega} \alpha R^2 = \mu^2  {\tilde R} + 6 \mu^2 (\Box \Omega - \partial^\mu \Omega \partial_\mu \Omega)  - \frac{\mu^4}{4 \alpha} 
\eeq
After further eliminating a total divergence, we see that the action appears now in the Einstein frame:
\begin{equation}
{\cal A} \; = \; \frac{1}{2}\int d^4x \sqrt{- {\tilde g}} \left( \mu^2 {\tilde R} - 6 \mu^2 \partial^\mu \Omega \partial_\mu \Omega - \frac{\mu^4}{4 \alpha} \right) \, .
\label{R2Einstein0}
\end{equation}
or equivalently
\beq
{\cal A} \; = \; \frac{1}{2}\int d^4x \sqrt{- {\tilde g}} \left( \mu^2 {\tilde R} - \frac{3 \mu^2}{2} \frac{\partial ^\mu \Phi \partial_\mu \Phi}{\Phi^2} - \frac{\mu^4}{4 \alpha} \right) \, .
\label{R2Einstein1}
\end{equation}
Finally, in order to write the scalar kinetic term in canonical form, we introduce $\phi \equiv \sqrt{6} \mu \Omega$,
leading to
\begin{equation}
{\cal A} \; = \; \frac{1}{2}\int d^4x \sqrt{- {\tilde g}} \left( \mu^2 {\tilde R} - \partial ^\mu \phi \partial_\mu \phi - \frac{\mu^4}{4 \alpha} \right) \, .
\label{R2Einstein2}
\end{equation}
We see now that the mass scale $\mu$ can be identified with the Planck scale:
Newton's constant $8 \pi G_N^2 = \kappa^2 = 1/\mu^2$.

Thus we have recovered the well-known result~\cite{old,KLT} that pure $R^2$ gravity is equivalent to the conventional Einstein-Hilbert
theory with a massless scalar field $\phi$ and a cosmological constant $\Lambda = \mu^4/8 \alpha$.
The dimensionless parameter $\alpha$ in (\ref{pureR2}) specifies the magnitude of $\Lambda$ in
Planck units, and we see that $\alpha \gg 1$ is required.  \\

\noindent
{\it The Starobinsky Model of Inflation}\\

Next we recall the Starobinsky model of inflation~\cite{Starobinsky}, which is derived by
adding to the pure $R^2$ action (\ref{pureR2}) the conventional linear Einstein-Hilbert term:
\begin{equation}
{\cal A} \; = \; \frac{1}{2 \kappa^2} \int d^4x \sqrt{-g} \left( R + {\tilde \alpha} R^2 \right) \, ,
\label{EH}
\end{equation}
where we have introduced the dimensionful constant ${\tilde \alpha} = \kappa^2 \alpha$. 
As is well known, after rewriting $\alpha R^2 \to 2 \alpha \Phi R - \alpha\Phi^2$ and making a conformal transformation
analogous to that in the pure $R^2$ case:
\begin{equation}
{\tilde g}_{\mu \nu}  \; = \; e^{2\Omega} g_{\mu \nu} \; = \; \left(1 + 2 {\tilde \alpha} \Phi \right) g_{\mu \nu} \, ,
\end{equation}
one finds
\begin{equation}
{\cal A} \; = \; \frac{1}{2 \kappa^2}  \int d^4x \sqrt{- {\tilde g}} \left[ {\tilde R} - \frac{6 {\tilde \alpha}^2}{(1 + 2 {\tilde \alpha}  \Phi)^2} \left(\partial^\mu \Phi \partial_\mu \Phi +  \frac{\Phi^2}{6 {\tilde \alpha}} \right) \right] \, .
\label{almostStaro}
\end{equation}
Setting $\kappa \phi \equiv \sqrt{3/2} \ln \left(1 +  2 {\tilde \alpha} \Phi \right)$, 
(\ref{EH}) may be written as follows in the Einstein frame:
\begin{equation}
{\cal A} \; = \;  \frac{1}{2 \kappa^2} \int d^4x \sqrt{- {\tilde g}} \left[ {\tilde R} -\kappa^2 \partial^\mu \phi \partial_\mu \phi - \frac{1}{4 {\tilde \alpha}} 
\left(1 - e^{-\sqrt{\frac{2}{3}}\kappa \phi }
 \right)^2 \right] \, .
\label{FullStaro}
\end{equation}
Thus one recovers the successful inflationary potential (\ref{Staro}) with ${\tilde \alpha} = 1/6 M^2$~\cite{Starobinsky}.
The scale invariance of the pure $R^2$ theory (\ref{pureR2})
is broken explicitly by the Einstein-Hilbert term in (\ref{EH}), and leads to an effective potential (\ref{Staro}) with a constant, scale-invariant asymptotic
limit that is approached exponentially at a rate controlled by the Planck scale $\kappa$.\\

\noindent
{\it Generalization with additional conformally-coupled fields}\\

With a view to the later comparison with generalized no-scale supergravity models~\cite{EKN}, we now consider adding to the $R^2$ action (\ref{EH}) $N-1$
additional complex fields $\phi^i$ with conformal couplings to $R$:
\begin{equation}
{\cal A} \; = \;   \frac{1}{2 \kappa^2} \int d^4x \sqrt{-g} \left[ \delta R + {\tilde \alpha} R^2 - 2 \kappa^2 \sum_{i=1}^{N-1} \left(\partial^\mu \phi^i \partial_\mu \phi^\dagger_i  + \frac{1}{3} |\phi^i|^2 R \right) \right] \, ,
\label{manyphi}
\end{equation}
where $\delta = 0$ corresponds to the $R^2$ theory and we allow $\delta = 1$ to describe the $R+ \alpha R^2$ Starobinsky model. 
As previously, we introduce a Lagrange multiplier field $\Phi$, and replace $ \alpha R^2$ in (\ref{manyphi})
by $2 \alpha \Phi R - \alpha \Phi^2$, as in (\ref{phiR2}). In order to transform to the Einstein frame, 
we must now rescale the metric by the modified conformal factor $\Omega$:
\begin{equation}
{\tilde g}_{\mu \nu}  \; = \; e^{2\Omega} g_{\mu \nu} \; = \; (\delta + 2 {\tilde \alpha} \Phi - \frac{\kappa^2}{3} \sum_{i = 1}^{N-1} |\phi^i |^2 ) g_{\mu \nu}  \, .
\label{newtilde}
\end{equation}
Thus we arrive at the following generalization of (\ref{R2Einstein0}):
\begin{equation}
\label{R2Einstein3}
\hspace{-1cm}
{\cal A} \; = \;   \frac{1}{2 \kappa^2} \int d^4x \sqrt{-{\tilde g}} \left[ {\tilde R} - 6 \partial^\mu \Omega \partial_\mu \Omega
\; - \; \sum_{i=1}^{N-1} \frac{ 2 \kappa^2 \partial^\mu \phi^i \partial_\mu \phi^\dagger_i }{\left(\delta + 2 {\tilde \alpha}  \Phi - \frac{\kappa^2}{3} 
\sum_{i = 1}^{N-1} |\phi^i |^2 \right)} -  \frac{ {\tilde \alpha}  \Phi^2}{\left(\delta + 2 {\tilde \alpha}  \Phi - \frac{\kappa^2}{3} \sum_{i = 1}^{N-1} |\phi^i |^2 \right)^2}  \right]
\end{equation}
that we compare below with the effective action of SU(N,1)/SU(N)$\times$U(1) no-scale supergravity~\cite{ENO}.\\

\noindent
{\it From $R^2$ Gravity to SU(1,1)/U(1) No-Scale Supergravity}\\

The pure $R^2$ supergravity model was constructed in~\cite{FKP}. Here we compare the action in (\ref{R2Einstein1}) and (\ref{R2Einstein2})
with that of the simplest SU(1,1)/U(1) no-scale supergravity model \cite{EKN1} (see also \cite{KLT}).

We recall that, in addition to the supergravity multiplet, which contributes the bosonic term ${\cal A} \; = \; \frac{1}{2 \kappa^2} \int d^4x \sqrt{-g} R$
to the effective action, the structure of the matter sector of a supergravity theory is characterized by the K\"ahler function
\beq
G = K + \ln |W|^2 \, ,
\eeq
where the K\"ahler potential $K$ is a hermitian function of the complex scalar fields, and
$W$ is the superpotential, which is a holomorphic function of these fields.
The simplest SU(1,1)/U(1) no-scale supergravity model can be written in terms of a single complex field $T$ with K\"ahler potential
\beq
K =  - 3 \ln \kappa \left(T + T^\dagger \right) \, ,
\label{su11}
\eeq
whose Lagrangian takes the very simple form
\beq
{\cal L} = -\frac{3}{(T+T^\dagger)^2 \kappa^2} \partial^\mu T \partial_\mu T^\dagger = - \frac{1}{12\kappa^2}(\partial_{\mu} K)^2 
- \frac34 e^{2K/3}|\partial_{\mu}(T-T^\dagger )|^2 \, ,
\label{boskin1}
\eeq
with vanishing potential.

In order to establish the correspondence with $R^2$ gravity, we 
consider a superpotential of the form \cite{EKN1,RS}
\beq
W = T^3 - \frac{\mu^3}{12 \alpha} \, ,
\label{T3-r}
\eeq
which generates a scalar potential of the form
\beq
V(T,T^\dagger) = \frac{\mu^4}{4 \alpha} \frac{T^2 + {T^\dagger}^2}{(T + {T^\dagger})^2} \, .
\label{V1}
\eeq
We restrict our attention initially to the real direction in field space~\footnote{As noted in \cite{RS}, this theory is unstable in the imaginary
$T - T^\dagger$ direction, but could be stabilized by some suitable mechanism such as quartic terms in the K\"ahler potential, as considered in~\cite{EKN,ENO7}.}.
With this restriction,  the last term in (\ref{boskin1}) can be discarded, and the 
effective potential (\ref{V1}) is constant along the real $T + T^\dagger$ direction, {\it \`a la} De Sitter.
Comparing with the action (\ref{R2Einstein2}), which is equivalent to the pure $R^2$ theory (\ref{pureR2}),
we see that there is a direct correspondence, and that we can identify 
\beq
 K = -6 \Omega \, .
 \label{OmegaK}
 \eeq
 Thus we have made the association $\delta + 2 {\tilde \alpha} \Phi = \kappa (T+ T^\dagger)$ and $\Omega = \frac12 \ln \kappa(T+T^\dagger)$.
It is striking that this correspondence is realized with a superpotential (\ref{T3-r})
that is a simple combination of trilinear and constant terms.

We note that the relation (\ref{OmegaK}) holds in general when one matches to supergravity any theory whose gravitational kinetic 
term can be written in the
form $\Phi R$, including $R^2$ gravity. The supergravity Lagrangian can be written as \cite{cremmer}
\beq
{\cal L}_{SG} = -\frac{1}{6} \Phi R - \frac{\partial^2 \Phi}{\partial \phi^i \partial \phi_j^*} (\partial_\mu \phi^i \partial^\mu \phi_j^*)  - \frac{1}{4\Phi}(\frac{\partial \Phi}{\partial \phi^i } \partial_\mu \phi^i - \frac{\partial \Phi}{\partial \phi_j^* }\partial_\mu \phi_j^*)^2+ \dots \, ,
\label{Lsg}
\eeq
where the $\dots$ represent terms containing gauge fields and fermions, and potential terms. Here, $\Phi$ is a real function of the scalar components
of chiral superfields. Upon transformation to the Einstein frame via a conformal transformation with 
$e^{2\Omega} = -\kappa^2 \Phi/3$, we recover the standard kinetic terms for supergravity with
\beq
{\cal L} = \frac{1}{2\kappa^2} {\tilde R} - \frac{1}{\kappa^2}K_i^j (\partial_\mu \phi^i)(\partial^\mu \phi_j^*)
\eeq
where $K_i \equiv \partial K/\partial \phi^i$ and $K^i \equiv \partial K/\partial \phi_i^*$ and we have the 
same relation (\ref{OmegaK}) between $K$ and $\Omega$. 
We further note that, since the pure $R^2$ gravitational theory contains no kinetic term for scalar fields, i.e., the middle term in Eq. (\ref{Lsg})
vanishes, the kinetic term of the scalar degree of freedom in the Einstein frame arises solely from the conformal
transformation \cite{EKN1}, and we can write $-\kappa \Phi/3 = T + T^\dagger$, i.e., $K=-3\ln \kappa (T + T^\dagger)$, without loss of generality,
thus pointing to the root of the $R^2$ conformal equivalence to no-scale supergravity.

 The correspondence between the kinetic terms for the conformal scalar field in the 
Starobinsky model (\ref{FullStaro}) and the no-scale field in (\ref{boskin1}), namely
$\delta + 2 {\tilde \alpha} \Phi = \kappa (T+ T^\dagger)$, was already noted in \cite{ENO7}.  This identification
reflects the partial invariance of both theories under the non-compact U(1) scale transformations: $t \; \to \; \alpha t$, which are included
in the SU(1,1) group of isometric transformations as dilations, though neither $K$ nor $W$ are themselves invariant,
and the analogous transformation for the scalar kinetic term in the Starobinsky model (\ref{FullStaro}).
The potential of the Starobinsky model or the corresponding no-scale SU(1,1)/U(1) model
is, however, not invariant under this rescaling of the corresponding scalar field, as this scale invariance is explicitly broken
by the Einstein-Hilbert term, which is linear in the curvature $R$, or by the superpotential, causing a deviation from pure De Sitter.

The kinetic term for the imaginary part of $T$ can also be accounted for if we extend the 
gravitational action to include an auxiliary field, $b_\mu$, coupled as follows in the Einstein frame
\beq
\Delta {\cal A} =  - \frac{1}{\kappa^2} \int d^4x \sqrt{-{\tilde g}}\, (\frac13 b_\mu b^\mu - b_\mu J^\mu)
\eeq
to a current $J_\mu$:
\beq
J_\mu = -2 (\Omega_T \partial_\mu T - \Omega_{T^*} \partial_\mu T^*) = \frac13 (K_T \partial_\mu T - K_{T^*} \partial_\mu T^*) \, .
\eeq
The field $b_\mu$ satisfies the equation of motion $b_\mu =  \frac32 J_\mu$, so that the action becomes
\beq
\Delta {\cal A} =  \frac{3}{4\kappa^2} \int d^4x \sqrt{-{\tilde g}} \, (J_\mu  J^\mu ) = -   \frac{3}{4\kappa^2}  \int d^4x \sqrt{-{\tilde g}} \, |J_\mu |^2 \, ,
\label{j2}
\eeq
which corresponds to the final term in (\ref{boskin1}), see also \cite{EKN1}.

Before generalizing the SU(1,1)/U(1) theory, we return to the question of flat potentials and the SU(1,1) invariance. 
It was argued in \cite{EKN1} that, in order to solve the hierarchy problem, the theory should have constant K\"ahler curvature,
${\cal R} = 2/3$ which is guaranteed by the choice of K\"ahler potential given in (\ref{su11}). As was also shown 
in \cite{EKN1}, the SU(1,1) invariance also allows, more generally, any space with K\"ahler curvature given by
${\cal R} = 2/3a$ which is obtained when (\ref{su11}) is generalized to 
\beq
K =  - 3 a \ln \kappa \left(T + T^\dagger \right) \, .
\label{sua1}
\eeq
This theory will also produce a flat potential \cite{EKN1,LN,RS} when i) $W(T) = 1$ leading to zero cosmological constant, or
ii) when $W(T) =  T^{3a/2}$ leading to an anti-De Sitter solution \cite{ENO7}, or iii) when $W(T) = T^{3a/2}(T^{-3\sqrt{a}/2}-T^{3\sqrt{a}/2})$. 
The latter corresponds to the choice (\ref{T3-r}) for $a=1$ and is stable for $a > 1$ \cite{RS}.
It can be used for single-field inflation as in the so-called $\alpha$-attractor models \cite{alpha}.
We note that this class of maximally-symmetric models can also be matched to the $R^2$ theory with 
$K = -6a\Omega$, with the special case of $a=1$ corresponding to the no-scale models we discuss here.\\

\noindent
{\it Generalization to SU(N,1)/SU(N)$\times$U(1) No-Scale Supergravity}\\

We now show that the generalization (\ref{manyphi}) of the $R^2$ theory with additional conformally-coupled fields
corresponds in a similar way to the SU(N,1)/SU(N)$\times$U(1)
no-scale supergravity model~\cite{EKN} with K\"ahler potential (\ref{noscale}).

In this model,
the relevant scalar-bosonic kinetic terms can (after some simple algebraic manipulations) be written as
\beq
\label{boskin}
- \frac{1}{12\kappa^2}(\partial_{\mu} K)^2- e^{K/3}|\partial_{\mu}\phi^i|^2 
-\frac34 e^{2K/3}| \partial_{\mu}(T-T^{\dagger})
+ \sum^{N-1}_1 \frac13\kappa(\phi_i^{*}\partial_{\mu}\phi^i-\phi^i\partial_{\mu}\phi^{*}_i)|^2 \, .
\eeq
 Comparing (\ref{boskin}) with (\ref{R2Einstein3}), we can make the same identification as in (\ref{OmegaK}),
 after identifying $\delta + 2 {\tilde \alpha} \Phi = \kappa(T + T^*)$.
 As in the SU(1,1)/U(1) case, we assume that the imaginary component of $T$ is stabilized
 as well as the imaginary parts of $\phi^i$,
 in which case the last term in (\ref{boskin}) can be discarded, and the correspondence to the kinetic terms in
 (\ref{R2Einstein3}) is direct.

 The remaining kinetic terms associated with the imaginary parts of the scalar fields
 can be mirrored by making the same addition to the action as in (\ref{j2}),
 with an extension of  the current to include the remaining $N-1$ fields: 
\beq
J_\mu =  \frac13 \sum_a (K_a \partial_\mu \Phi^a - K^a \partial_\mu \Phi_a^*) \, ,
\label{j}
\eeq
where the index $a$ runs over the fields $T$ and the $\phi^i$. The current-current interaction in (\ref{j2})
corresponds to the final term in (\ref{boskin}) when the current is defined as in (\ref{j}).

In order to complete the correspondence with the generalized $R^2$ gravity theory (\ref{manyphi}),
we must introduce an effective scalar potential term corresponding to the last term in (\ref{R2Einstein3}).
This is easily done by including the same superpotential term as in (\ref{T3-r}), which yields the scalar potential
 \beq
 V(T,\phi) = \frac{\mu^2}{4 \alpha} \frac{T^2 + {T^\dagger}^2}{(\kappa(T + {T^\dagger})- \frac{\kappa^2}{3} \sum_{i = 1}^{N-1} |\phi^i |^2)^2} \, .
 \label{VTphi}
 \eeq
This reproduces the last term in Eq. (\ref{R2Einstein3}) for $\delta = 0$ when we restrict our attention to the real direction in $T$,
as per our previous assumption that the imaginary direction in $T$ is stabilized.

To summarize this part of our paper:  he conformal factor that transforms the generalized
 scale-invariant $R^2$ theory with multiple conformally-coupled scalar fields (\ref{manyphi}) to the Einstein frame is identical with the K\"ahler
 potential of SU(N,1)/SU(N)$\times$U(1) no-scale supergravity.
 The kinetic term for $\Omega$ in (\ref{R2Einstein3}) matches exactly the first term in the no-scale scalar kinetic term (\ref{boskin}),
and the second term in (\ref{boskin}) also exactly matches the kinetic term for the $\phi^i$ in (\ref{R2Einstein3}).
When we restrict our attention to the real parts of the complex fields $T$ and $\phi^i$, the last term in (\ref{boskin})
vanishes, and the identification is complete. This final term can also be mirrored by introducing into the gravity theory
a suitable interaction of the current (\ref{j}).\\

\noindent
{\it Introducing a Starobinsky-like Inflationary Potential}\\

We now discuss how a potential ${\hat V}$ for the fields $\phi^i$ may be introduced into the multi-field $R^2$ action (\ref{manyphi}), i.e.,
\beq
\Delta {\cal A} = - \frac{1}{2 \kappa^2} \int d^4x \sqrt{-g} 2 \kappa^2 {\hat V}(\Phi, \phi^i) \, ,
\label{potential}
\eeq
which corresponds to a term of the form
\beq
\Delta {\cal A} = - \frac{1}{2 \kappa^2} \int d^4x \sqrt{-{\tilde g}} \frac{2\kappa^2{\hat V}(\Phi, \phi^i)}{(\delta + 2 {\tilde \alpha} \Phi - \frac{\kappa^2}{3} \sum_{i = 1}^{N-1} |\phi^i |^2)^2}
\label{delA}
\eeq
 in the Einstein frame. 
 
 We recall that the effective potential in no-scale supergravity takes the following form for general $W$:
 \beq
 V = \frac{\hat V}{(\kappa(T + {T^\dagger})- \frac{\kappa^2}{3} \sum_{i = 1}^{N-1} |\phi^i |^2)^2} 
 \eeq
 where
  \beq
{\hat V} \; \equiv \; \sum_1^{N-1} \left| \frac{\partial W}{\partial \phi^i} \right|^2  +\frac{\kappa}{3} (T+T^*) |W_T|^2 +
\frac{\kappa^2}{3} \left(W_T (\sum_1^{N-1}\phi_i^* W_{\phi_i}^* - 3 W^*) + {\rm h.c.}  \right) \, .
\label{effV}
\end{equation}
The following specific, separable form for $W$:
\beq
W = T^3  - \frac{\mu^3}{12 \alpha}  + f(\phi^i) \, ,
\label{minW}
\eeq 
yields the following form for the scalar potential ${\hat V}$:
 \beq
 {\hat V}(T,\phi^i) = (\kappa^2(\phi_i f_{\phi_i} - 3 f) {T^\dagger}^2  + h.c. ) + |f_{\phi^i}|^2 + \kappa\frac{\mu^3}{4 \alpha}(T^2 + {T^\dagger}^2)\, .
 \label{vhatgen}
 \eeq
 We note that the last term in  (\ref{vhatgen}) is precisely that in (\ref{VTphi}), and does not contribute to the
 construction of ${\hat V}(\phi,\phi^i)$ in (\ref{delA}).
 
In order to realize inflation in this framework,
we restrict our attention, for simplicity, to a single matter field $\phi^1$ in addition to the modulus $T$
 in the no-scale picture. This corresponds to a non-compact SU(2,1)/SU(2)$\times$U(1) coset structure,
 which we have argued previously is the minimal structure required to construct a suitable inflationary model~\cite{ENO}. 
 We consider two forms of the superpotential which can accommodate the
 Starobinsky model of inflation.  
 
 The first is a Wess-Zumino model  in which the inflaton is identified with $\phi^1$. It is described by the $\phi^1$-dependent superpotential~\cite{ENO}
 \beq
 W^{WZ} \; = \; M \left[\frac{{\phi^1}^2}{2} - \frac{{\kappa \phi^1}^3}{3 \sqrt{3}} \right] \, .
 \label{WZmodel}
\eeq
 As discussed in~\cite{ENO}, if we assume that $T$ is constrained by Planck-scale dynamics to have the specific value 
$\kappa T= 1/2$~\footnote{The choice of this value is for
 illustration: other choices yield similar results when combined with the corresponding change in (\ref{WZmodel}).},  the resulting no-scale model
 yields the Starobinsky potential (\ref{Staro}), as we now show. 
 
 Restricting to real fields as discussed previously,  we can match this theory with the potential
 \beq
 {\hat V} = {M^2} {\phi^1}^2 \left(1-\kappa \phi^1/\sqrt{3} \right)^2
 -\frac{1}{2\kappa^2} {\tilde \alpha} \Phi^2  \, ,
 \label{vhatwz}
 \eeq
 where the last term in (\ref{vhatwz}) is needed to cancel the last term in (\ref{R2Einstein3}). Then,
 assuming that the value of $\Phi$ is fixed: $\delta + 2{\tilde \alpha} \Phi = 1$, 
and combining with (\ref{R2Einstein3}), we find the following scalar potential for $\phi^1$:
\beq
V(\phi^1) = \frac{{\hat V} + \frac{(1-\delta)^2}{8{\tilde \alpha} \kappa^2}}{(1 - \frac{\kappa^2}{3} |\phi^1|^2)^2} = \frac{M^2 {\phi^1}^2 \left(1-\kappa \phi^1/\sqrt{3} \right)^2}{(1 - \frac{\kappa^2}{3} |\phi^1|^2)^2}
\eeq 
when one recalls that ${\tilde \alpha} = 1/6M^2$, and remembers that the last term in (\ref{vhatwz}) cancels.
Finally, making the transformation
\beq
\phi^1 = \sqrt{3} \tanh (\phi/\sqrt{6}) \, ,
\eeq
 one recovers the standard form of the Starobinsky potential (\ref{Staro}) as a function of this $\phi$ field.
 
The second route to a Starobinsky-like model of inflation is the Cecotti model~\cite{Cecotti}:
 \beq
 W^C \; = \sqrt{3} M \phi^1 (T - 1/2) \, ,
 \label{WC}
 \eeq
 where the inflaton is identified with $T$.
This offers a simpler realization of inflation, since it does not require any additional superpotential term as in Eq.(\ref{T3-r}). 
In terms of $\Phi$ and $\phi^1$, the theory is equivalent (when fields are again taken to be real) to
\beq
2 \kappa^2 {\tilde \alpha} {\hat V} = \frac14 -  \frac{\delta + 2{\tilde \alpha} \Phi}{2} + 
\frac{\delta^2 + 4 \delta {\tilde \alpha} \Phi}{4} + \frac23 \kappa^2 {\phi^1}^2 ( 1- \frac{\delta + 2 {\tilde \alpha} \Phi}{2} ) \, .
\eeq
When $\phi^1$ is constrained to have vanishing expectation value and 
when combined with (\ref{R2Einstein3}), this yields the following scalar potential for $\Phi$:
\beq
V(\phi) =  \frac{{\hat V} + ({\tilde \alpha} \Phi^2/2 \kappa^2)}{(\delta + 2 {\tilde \alpha} \Phi)^2} = \frac{3 M^2}{16 \kappa^2 t^2} -\frac{3M^2}{4 \kappa^2 t}
+  \frac{3M^2}{4 \kappa^2} \, ,
\eeq
where we have defined $2 t = \delta + 2 {\tilde \alpha} \Phi$. 
Making the transformation $\kappa \phi \equiv \sqrt{6} \Omega = (\sqrt{6}/2) \ln 2 t $
we recover once again the Starobinsky potential as a function of this $\phi$ field.\\

\noindent
{\it Summary}\\

We have explored in this paper the connection between $R^2$ gravity and minimal SU(1,1)/U(1)
no-scale supergravity \cite{ENO}. The supersymmetric completion of pure $R^2$ was considered in \cite{FKP}, and 
we have shown that by an extension of the theory to include many
conformally-coupled scalar fields can be matched to more general SU(N,1)/SU(N)$\times$U(1)
no-scale supergravity theories (see also \cite{KLT}). 
In the $R^2$ frame, we are able to perform a conformal transformation $e^\Omega$ to the Einstein frame
which introduces a dynamical real scalar degree of freedom, $\Omega$.  In the pure $R^2$ theory, the field is massless
and there is a non-zero cosmological constant characterized by the coupling of the $R^2$ term in the action.
In either the $R+R^2$ theory, or one with conformally coupled scalar fields, the scalar potential is non-trivial
and possesses a second-order pole, whereas the kinetic terms of the conformally-coupled scalars contain a first-order pole. 
The presence of this pole leads to the asymptotically-flat feature at large field values that is characteristic of the Starobinsky model~\cite{Starobinsky}.
In the no-scale supergravity theory, due to the logarithmic structure of the K\"ahler potential
and the definition of the kinetic and potential terms in terms of derivatives of $K$, these terms 
(in the real directions) possess exactly the same pole structure,
leading to the asymptotical flatness so useful for inflation. 

We have shown that the parts of the action involving the real components of the scalar fields are identical
in $R^2$ gravity and no-scale supergravity, and we have shown how this correspondence can be
extended to the imaginary components by adding a suitable current-current interaction to the $R^2$
gravity theory. Our analysis deepens understanding of the connection between no-scale supergravity
and scale-invariant extensions of Einstein's theory of gravity. Our interest in this connection was triggered 
by the observational success~\cite{planck15} of the Starobinsky model of inflation~\cite{Starobinsky}, and we have reviewed briefly above
two examples how a Starobinsky-like inflationary potential can emerge in simple ways from
SU(2,1)/SU(2)$\times$U(1) no-scale supergravity theory. We think that this is the most promising
avenue for constructing eventually a complete theory of everything below the Planck scale,
connecting inflationary model-building to accessible physics beyond the Standard Model~\cite{ENO8,EGNO4,EGNO5,EGNNO,EGNNO2}.

\section*{Acknowledgements}

We would like to thank M. Voloshin for helpful conversations. 
The work of J.E. was supported in part by STFC (UK) via the research grant ST/L000326/1
and in part by the Estonian Research Council via a Mobilitas Pluss grant. The work of D.V.N. was supported in part by the DOE
grant DE-FG02-13ER42020 and in part by the Alexander~S.~Onassis Public
Benefit Foundation. The work of K.A.O. was supported in part by
DOE grant DE-SC0011842 at the University of Minnesota.


\newpage
\begin{thebibliography}{9}

\bibitem{Starobinsky}
A.~A.~Starobinsky,
  Phys.\ Lett.\ B {\bf 91}, 99 (1980).
  
  \bibitem{MukhChib}
    V.~F.~Mukhanov and G.~V.~Chibisov,
  JETP Lett.\  {\bf 33}, 532 (1981)
  [Pisma Zh.\ Eksp.\ Teor.\ Fiz.\  {\bf 33}, 549 (1981)].
 
  \bibitem{planck15}
 P.~A.~R.~Ade {\it et al.} [Planck Collaboration],
  Astron.\ Astrophys.\  {\bf 594}, A13 (2016)
  [arXiv:1502.01589 [astro-ph.CO]].
  P.~A.~R.~Ade {\it et al.} [Planck Collaboration],
  Astron.\ Astrophys.\  {\bf 594}, A20 (2016)
  [arXiv:1502.02114 [astro-ph.CO]].
  
 \bibitem{Cries}
J.~R.~Ellis, D.~V.~Nanopoulos, K.~A.~Olive and K.~Tamvakis,
  Phys.\ Lett.\  {\bf 118B} (1982) 335;
  J.~R.~Ellis, D.~V.~Nanopoulos, K.~A.~Olive and K.~Tamvakis,
  Nucl.\ Phys.\ B {\bf 221} (1983) 52;
  K.~Nakayama and F.~Takahashi,
  JCAP {\bf 1110}, 033 (2011)
  [arXiv:1108.0070 [hep-ph]].
  
  \bibitem{nost}
  D.~V.~Nanopoulos, K.~A.~Olive, M.~Srednicki and K.~Tamvakis,
  Phys.\ Lett.\ B {\bf 123}, 41 (1983);
    R.~Holman, P.~Ramond and G.~G.~Ross,
  Phys.\ Lett.\ B {\bf 137}, 343 (1984);
  A.~B.~Goncharov and A.~D.~Linde,
  Phys.\ Lett.\ B {\bf 139}, 27 (1984).
  
\bibitem{GL}
   A.~S.~Goncharov and A.~D.~Linde,
  Class.\ Quant.\ Grav.\  {\bf 1},  L75 (1984).

\bibitem{KQ}
C.~Kounnas and M.~Quiros,
  Phys.\ Lett.\ B {\bf 151}, 189 (1985).

  
 \bibitem{EENOS}
J.~R.~Ellis, K.~Enqvist, D.~V.~Nanopoulos, K.~A.~Olive and M.~Srednicki,
  Phys.\ Lett.\  {\bf 152B} (1985) 175
   Erratum: [Phys.\ Lett.\  {\bf 156B} (1985) 452].
  
   \bibitem{otherns}
  K.~Enqvist, D.~V.~Nanopoulos and M.~Quiros,
  Phys.\ Lett.\ B {\bf 159}, 249 (1985);
        P.~Bin\'etruy and M.~K.~Gaillard,
  Phys.\ Rev.\ D {\bf 34}, 3069 (1986);
   H.~Murayama, H.~Suzuki, T.~Yanagida and J.~Yokoyama,
  Phys.\ Rev.\  D {\bf 50}, 2356 (1994)
  [arXiv:hep-ph/9311326];
    S.~C.~Davis and M.~Postma,
  JCAP {\bf 0803}, 015 (2008)
  [arXiv:0801.4696 [hep-ph]];
    S.~Antusch, M.~Bastero-Gil, K.~Dutta, S.~F.~King and P.~M.~Kostka,
  JCAP {\bf 0901}, 040 (2009)
  [arXiv:0808.2425 [hep-ph]];
  S.~Antusch, M.~Bastero-Gil, K.~Dutta, S.~F.~King and P.~M.~Kostka,
  Phys.\ Lett.\  B {\bf 679}, 428 (2009)
  [arXiv:0905.0905 [hep-th]];
 R.~Kallosh and A.~Linde,
  JCAP {\bf 1011}, 011 (2010)
  [arXiv:1008.3375 [hep-th]];
    R.~Kallosh, A.~Linde and T.~Rube,
  Phys.\ Rev.\  D {\bf 83}, 043507 (2011)
  [arXiv:1011.5945 [hep-th]];
  S.~Antusch, K.~Dutta, J.~Erdmenger and S.~Halter,
  JHEP {\bf 1104} (2011) 065
  [arXiv:1102.0093 [hep-th]];
  R.~Kallosh, A.~Linde, K.~A.~Olive and T.~Rube,
  Phys.\ Rev.\ D {\bf 84}, 083519 (2011)
  [arXiv:1106.6025 [hep-th]];
   T.~Li, Z.~Li and D.~V.~Nanopoulos,
  JCAP {\bf 1402}, 028 (2014)
  [arXiv:1311.6770 [hep-ph]];
 W.~Buchmuller, C.~Wieck and M.~W.~Winkler,
  Phys.\ Lett.\ B {\bf 736}, 237 (2014)
  [arXiv:1404.2275 [hep-th]].

  
    \bibitem{no-scale}
E.~Cremmer, S.~Ferrara, C.~Kounnas and D.~V.~Nanopoulos,
  Phys.\ Lett.\ B {\bf 133} (1983) 61;
  J.~R.~Ellis, A.~B.~Lahanas, D.~V.~Nanopoulos and K.~Tamvakis,
  Phys.\ Lett.\ B {\bf 134} (1984) 429.
  
  \bibitem{EKN}
J.~R.~Ellis, C.~Kounnas and D.~V.~Nanopoulos,
  Nucl.\ Phys.\ B {\bf 247} (1984) 373.
  
    \bibitem{LN}
  A.~B.~Lahanas and D.~V.~Nanopoulos,
  Phys.\ Rept.\  {\bf 145} (1987) 1.
  
  \bibitem{GMO}
M.~K.~Gaillard, H.~Murayama and K.~A.~Olive,
  Phys.\ Lett.\ B {\bf 355} (1995) 71
  [hep-ph/9504307].
  
  \bibitem{drt}
  M.~Dine, L.~Randall and S.~D.~Thomas,
  Phys.\ Rev.\ Lett.\  {\bf 75}, 398 (1995)
  [hep-ph/9503303].
  
  \bibitem{deln}
  G.~A.~Diamandis, J.~R.~Ellis, A.~B.~Lahanas and D.~V.~Nanopoulos,
  Phys.\ Lett.\ B {\bf 173}, 303 (1986).

\bibitem{EKN1}
J.~R.~Ellis, C.~Kounnas and D.~V.~Nanopoulos,
  Nucl.\ Phys.\ B {\bf 241}, 406 (1984).



\bibitem{ENO}
   J.~Ellis, D.~V.~Nanopoulos and K.~A.~Olive,
  Phys.\ Rev.\ Lett.\  {\bf 111} (2013) 111301 
  [arXiv:1305.1247 [hep-th]].

     \bibitem{Cecotti}
S.~Cecotti,
  Phys.\ Lett.\ B {\bf 190} (1987) 86.
  
\bibitem{DLT}
G.~D.~Diamandis, A.~B.~Lahanas and K.~Tamvakis,
  Phys.\ Rev.\ D {\bf 92} (2015) no.10,  105023
  [arXiv:1509.01065 [hep-th]];
   G.~A.~Diamandis, B.~C.~Georgalas, K.~Kaskavelis, A.~B.~Lahanas and G.~Pavlopoulos,
  Phys.\ Rev.\ D {\bf 96}, no. 4, 044033 (2017)
  [arXiv:1704.07617 [hep-th]].
 
  
  
\bibitem{WhittStelle}
K.~S.~Stelle,
  Gen.\ Rel.\ Grav.\  {\bf 9} (1978) 353;
  B.~Whitt,
  Phys.\ Lett.\  {\bf 145B} (1984) 176.
  
\bibitem{KLT}
  C.~Kounnas, D.~L\"{u}st and N.~Toumbas,
  Fortsch.\ Phys.\  {\bf 63}, 12 (2015)
  [arXiv:1409.7076 [hep-th]].

\bibitem{ENO7}
   J.~Ellis, D.~V.~Nanopoulos and K.~A.~Olive,
  JCAP {\bf 1310} (2013) 009
 [arXiv:1307.3537 [hep-th]].

  \bibitem{EGNO4} 
  J.~Ellis, M.~A.~G.~Garc{\' i}a, D.~V.~Nanopoulos and K.~A.~Olive,
  JCAP {\bf 1510}, 003 (2015)
  [arXiv:1503.08867 [hep-ph]].
  
  \bibitem{FeKR}
  S.~Ferrara, A.~Kehagias and A.~Riotto,
  Fortsch.\ Phys.\  {\bf 62}, 573 (2014)
  [arXiv:1403.5531 [hep-th]];
S.~Ferrara, A.~Kehagias and A.~Riotto,
  Fortsch.\ Phys.\  {\bf 63}, 2 (2015)
  [arXiv:1405.2353 [hep-th]];
 R.~Kallosh, A.~Linde, B.~Vercnocke and W.~Chemissany,
  JCAP {\bf 1407}, 053 (2014)
  [arXiv:1403.7189 [hep-th]];
 K.~Hamaguchi, T.~Moroi and T.~Terada,
  Phys.\ Lett.\ B {\bf 733}, 305 (2014)
  [arXiv:1403.7521 [hep-ph]];
 J.~Ellis, M.~A.~G.~Garc\'ia, D.~V.~Nanopoulos and K.~A.~Olive,
  JCAP {\bf 1405}, 037 (2014)
  [arXiv:1403.7518 [hep-ph]];
J.~Ellis, M.~A.~G.~Garc{\' i}a, D.~V.~Nanopoulos and K.~A.~Olive,
  JCAP {\bf 1408}, 044 (2014)
  [arXiv:1405.0271 [hep-ph]].



\bibitem{others}
R.~Kallosh and A.~Linde,
  JCAP {\bf 1306} (2013) 028
  [arXiv:1306.3214 [hep-th]];
   T.~Li, Z.~Li and D.~V.~Nanopoulos,
  JCAP {\bf 1404}, 018 (2014)
  [arXiv:1310.3331 [hep-ph]];
  J.~Ellis, D.~V.~Nanopoulos and K.~A.~Olive,
  Phys.\ Rev.\ D {\bf 89} (2014) 4,  043502
  [arXiv:1310.4770 [hep-ph]];
  C.~P.~Burgess, M.~Cicoli and F.~Quevedo,
  JCAP {\bf 1311} (2013) 003
  [arXiv:1306.3512 [hep-th]];
F.~Farakos, A.~Kehagias and A.~Riotto,
  Nucl.\ Phys.\ B {\bf 876}, 187 (2013)
  [arXiv:1307.1137 [hep-th]];
  S.~Ferrara, R.~Kallosh, A.~Linde and M.~Porrati,
  Phys.\ Rev.\ D {\bf 88} (2013) 8,  085038
  [arXiv:1307.7696 [hep-th]];
  W.~Buchm\"uller, V.~Domcke and C.~Wieck,
  Phys.\ Lett.\ B {\bf 730}, 155 (2014)
  [arXiv:1309.3122 [hep-th]];
  C.~Pallis,
  JCAP {\bf 1404}, 024 (2014)
  [arXiv:1312.3623 [hep-ph]];
   C.~Pallis,
  JCAP {\bf 1408}, 057 (2014)
  [arXiv:1403.5486 [hep-ph]];
I.~Antoniadis, E.~Dudas, S.~Ferrara and A.~Sagnotti,
  Phys.\ Lett.\ B {\bf 733}, 32 (2014)
  [arXiv:1403.3269 [hep-th]];
  T.~Li, Z.~Li and D.~V.~Nanopoulos,
  Eur.\ Phys.\ J.\ C {\bf 75}, no. 2, 55 (2015)
  [arXiv:1405.0197 [hep-th]];
     W.~Buchmuller, E.~Dudas, L.~Heurtier and C.~Wieck,
  JHEP {\bf 1409}, 053 (2014)
  [arXiv:1407.0253 [hep-th]];
    J.~Ellis, M.~A.~G.~Garc\'ia, D.~V.~Nanopoulos and K.~A.~Olive,
  JCAP {\bf 1501}, no. 01, 010 (2015)
  [arXiv:1409.8197 [hep-ph]];
   T.~Terada, Y.~Watanabe, Y.~Yamada and J.~Yokoyama,
  JHEP {\bf 1502}, 105 (2015)
  [arXiv:1411.6746 [hep-ph]];
  W.~Buchmuller, E.~Dudas, L.~Heurtier, A.~Westphal, C.~Wieck and M.~W.~Winkler,
  JHEP {\bf 1504}, 058 (2015)
  [arXiv:1501.05812 [hep-th]];
 A.~B.~Lahanas and K.~Tamvakis,
  Phys.\ Rev.\ D {\bf 91}, no. 8, 085001 (2015)
  [arXiv:1501.06547 [hep-th]];

\bibitem{FKP}
 S.~Ferrara, A.~Kehagias and M.~Porrati,
  JHEP {\bf 1508} (2015) 001
  doi:10.1007/JHEP08(2015)001
  [arXiv:1506.01566 [hep-th]].
  
\bibitem{Moreothers}
I.~Dalianis and F.~Farakos,
  JCAP {\bf 1507}, no. 07, 044 (2015)
  [arXiv:1502.01246 [gr-qc]].
  I.~Garg and S.~Mohanty,
  Phys.\ Lett.\ B {\bf 751}, 7 (2015)
  [arXiv:1504.07725 [hep-ph]];
  J.~Ellis, M.~A.~G.~Garc{\' i}a, D.~V.~Nanopoulos and K.~A.~Olive,
  JCAP {\bf 1507}, no. 07, 050 (2015)
  [arXiv:1505.06986 [hep-ph]];
  E.~Dudas and C.~Wieck,
  JHEP {\bf 1510}, 062 (2015)
  [arXiv:1506.01253 [hep-th]];
M.~Scalisi,
  JHEP {\bf 1512}, 134 (2015)
  [arXiv:1506.01368 [hep-th]];
   [arXiv:1506.01566 [hep-th]];
  J.~Ellis, M.~A.~G.~Garc{\' i}a, D.~V.~Nanopoulos and K.~A.~Olive,
  Class.\ Quant.\ Grav.\  {\bf 33}, no. 9, 094001 (2016)
  [arXiv:1507.02308 [hep-ph]];
  A.~Addazi and M.~Y.~Khlopov,
  Phys.\ Lett.\ B {\bf 766}, 17 (2017)
  [arXiv:1612.06417 [gr-qc]];
   C.~Pallis and N.~Toumbas,
 Adv.\ High Energy Phys.\  {\bf 2017}, 6759267 (2017)
  [arXiv:1612.09202 [hep-ph]];
  M.~C.~Romao and S.~F.~King,
  JHEP {\bf 1707}, 033 (2017)
  [arXiv:1703.08333 [hep-ph]];
  T.~Kobayashi, O.~Seto and T.~H.~Tatsuishi,
  arXiv:1703.09960 [hep-th];
 E.~Dudas, T.~Gherghetta, Y.~Mambrini and K.~A.~Olive,
  arXiv:1710.07341 [hep-ph];
  I.~Garg and S.~Mohanty,
  arXiv:1711.01979 [hep-ph].
  
  \bibitem{RS}
  D.~Roest and M.~Scalisi,
  Phys.\ Rev.\ D {\bf 92}, 043525 (2015)
  [arXiv:1503.07909 [hep-th]].

  
    \bibitem{EGNNO}
J.~Ellis, M.~A.~G.~Garc{\' i}a, N.~Nagata, D.~V.~Nanopoulos and K.~A.~Olive,
  JCAP {\bf 1611} (2016) no.11,  018
  [arXiv:1609.05849 [hep-ph]].
  
  \bibitem{EGNNO2}
  J.~Ellis, M.~A.~G.~Garc{\' i}a, N.~Nagata, D.~V.~Nanopoulos and K.~A.~Olive,
  JCAP {\bf 1707} (2017) no.07,  006
  [arXiv:1704.07331 [hep-ph]].

  
\bibitem{old}
  H.~Buchdahl, Nuovo Cimento {\bf 23}, 141 (1962);
G.~V.~Bicknell, J. Phys. A {\bf 7}, 1061 (1974);
H.~Buchdahl, J. Phys. A {\bf 11}, 871 (1978).

\bibitem{cremmer}
E.~Cremmer, B.~Julia, J.~Scherk, S.~Ferrara, L.~Girardello and P.~van Nieuwenhuizen,
  Nucl.\ Phys.\ B {\bf 147}, 105 (1979);
E.~Cremmer, S.~Ferrara, L.~Girardello and A.~Van Proeyen,
  Nucl.\ Phys.\ B {\bf 212}, 413 (1983).
  doi:10.1016/0550-3213(83)90679-X

  
  \bibitem{alpha}
  R.~Kallosh, A.~Linde and D.~Roest,
  JHEP {\bf 1311}, 198 (2013)
  [arXiv:1311.0472 [hep-th]];
  R.~Kallosh, A.~Linde and D.~Roest,
  JHEP {\bf 1408}, 052 (2014)
  [arXiv:1405.3646 [hep-th]].
  

 \bibitem{ENO8}
      J.~Ellis, D.~V.~Nanopoulos and K.~A.~Olive,
  Phys.\ Rev.\ D {\bf 89}, 043502 (2014)
  [arXiv:1310.4770 [hep-ph]].
  
     \bibitem{EGNO5}
J.~Ellis, M.~A.~G.~Garc{\' i}a, D.~V.~Nanopoulos and K.~A.~Olive,
  JCAP {\bf 1507}, 050 (2015)
  [arXiv:1505.06986 [hep-ph]].
  
 
  
\end{thebibliography}
\end{document}